\documentclass[graybox, nosecnum]{svmult}

\usepackage{mathptmx}       
\usepackage{helvet}         
\usepackage{courier}        
\usepackage{type1cm}        
\usepackage{makeidx}         
\usepackage{graphicx}        
\usepackage{multicol}        
\usepackage[bottom]{footmisc}
\usepackage{hyperref}        
\usepackage{soul}            
\hypersetup{colorlinks=true,urlcolor=blue}
\usepackage[square,numbers]{natbib}
\makeindex             

\usepackage{fullpage}
\usepackage{amsmath}
\usepackage{amssymb}
\usepackage[utf8]{inputenc}
\usepackage[english]{babel}
\usepackage{enumitem}
\usepackage[dvipsnames]{xcolor}
\usepackage{graphics}
\usepackage{wrapfig}
\usepackage{caption}

\newcommand{\myemail}{wenbinlu@berkeley.edu}

\def\apj{{ ApJ}}
\def\apjl{{ApJL}}
\def\apjs{{ ApJS}}

\def\aap{{ A\&A}}

\def\mnras{{ MNRAS}}
\def\araa{{ ARA\&A}}

\def\nat {{ Nature}}

\def\ssr{{ Space Sci. Rev.}}

\def\prd{{ Phys. Rev. D}}

\def\nar{{New Astronomy Reviews}}

\def\mr{\mathrm}
\def\mc{\mathcal}
\def\d{\mr{d}}
\def\msun{M_\odot}
\def\msunyr{{M_\odot\mr{\,yr^{-1}}}}
\def\kB{k_{\rm B}}
\def\rsun{R_\odot}
\def\md{M_{\rm d}}
\def\Md{M_{\rm d}}
\def\rT{r_{\rm T}}
\def\rg{r_{\rm g}}
\def\rd{r_{\rm d}}
\def\rp{r_{\rm p}}
\def\rc{r_{\rm c}}
\def\rtrin{r_{\rm tr, in}}

\def\rtrfb{r_{\rm tr, fb}}
\def\rsph{r_{\rm sph}}
\def\mp{m_{\rm p}}

\def\omgK{\Omega_{\rm K}}
\def\dotMw{\dot{M}_{\rm w}}
\def\dotMd{\dot{M}_{\rm d}}
\def\dotMfb{\dot{M}_{\rm fb}}
\def\dotMbh{\dot{M}_{\rm bh}}
\def\dotMacc{\dot{M}_{\rm acc}}
\def\dotMEdd{\dot{M}_{\rm Edd}}

\def\fsh{f_{\rm sh}}
\def\fw{f_{\rm w}}
\def\Be{B_{\rm e}}

\def\tvis{t_{\rm vis}}
\def\nuvis{\nu_{\rm vis}}

\newcommand{\lrb}[1]{\left({#1}\right)}
\newcommand{\lrsb}[1]{\left[{#1}\right]}

\begin{document}
\title*{Review: Accretion disk evolution in tidal disruption events}

\author{Wenbin Lu}
\institute{Wenbin Lu \at Departments of Astronomy and Theoretical Astrophysics Center, UC Berkeley, Berkeley, CA 94720, USA\\
Department of Astrophysical Sciences, Princeton University, Princeton, NJ 08544, USA\\
\email{\myemail}}
%

\maketitle

\abstract{This is a brief review of the recent progress in understanding the evolution of the accretion disks in tidal disruption events (TDEs). Special attention is paid to (1) thermal-viscous instability that causes the disk to transition from a thick state to a thin one, and back and forth, (2) interactions between the fallback material and existing disk. Challenges to the current model from late-time X-ray observations are highlighted and possible solutions are discussed.
}
\section{Keywords} 
tidal disruption events --- hydrodynamics --- accretion disks --- relativistic jets




\section{1. Introduction}
Tidal disruption of a star occurs when it approaches a supermassive black hole (BH) within a distance of the order $\rT = R_*(M/M_*)^{1/3}$, for stellar mass $M_*$ and radius $R_*$, and BH mass $M$. These tidal disruption events (TDEs) generate a sudden burst of accretion onto the BH which powers bright electromagnetic emission \citep[see][for a comprehensive review of the observations and some basic theoretical ideas]{gezari21_TDE_review}. They provide an ideal lab to study the physics of accretion disks and relativistic jets, because the stellar debris falls back towards the BH and joins the accretion disk at a rate which starts at a super-Eddington level and then declines over time in a more-or-less known manner \citep{rees88_TDEs, phinney89_fallback_rate, evans89_fallback_rate, lodato09_fallback_rate, guillochon13_fallback_rate, cheng14_relativistic_disruptions, coughlin19_partial_TDE_fallback, ryu20_GR_TDE, law-smith20_STAR_library}, see \citet{rossi20_disruption_process_review} for a review of the stellar disruption phase. However, the hydrodynamics of the disk formation phase --- how the stellar debris initially in the form of a high-eccentric stream dissipates its orbital energy and forms a quasi-circular disk ---  remains uncertain \citep[see][for a recent review]{bonnerot21_disk_formation_review}. The key numerical challenge is to capture the hydrodynamic evolution of the very thin, supersonic (Mach number $\gtrsim 10^3$) fallback stream near the pericenter passage, including the ``nozzle shock'' which compresses the stream vertically into a thin sheet \citep{kochanek94_stream_evolution, guillochon14_GR_dissipation, bonnerot22_nozzle_shock}.



Nevertheless, it is widely believed that orbital circularization of the fallback material is initialized by the self-intersection of the debris stream caused by general relativistic (GR) apsidal precession \citep{guillochon14_GR_dissipation, shiokawa15_disk_formation, dai15_intersection_radius, bonnerot16_disk_formation, kimitake16_circularization, sadowski16_deep_TDE, jiang16_self-intersection, liptai19_GRSPH, lu20_self_intersection}. Afterwards, the gas expanding from the intersection point, as a result of a broad distribution of angular momentum, experiences numerous secondary shocks and is expected to quickly form an accretion disk \citep{bonnerot20_disk_formation, bonnerot21_disk_formation, curd21_stream_injection, andalman22_self_intersection}. During this process, a large fraction of the energy dissipated by secondary shocks and viscous processes may be reprocessed by the optically thick gas at large distances (of $10^{14}$--$10^{15}\rm\,cm$) into the optical/UV bands \citep{piran15_disk_formation, metzger16_reprocessing, roth16_reprocessing, dai18_unified_model, bonnerot21_disk_formation}. The radiative signatures of this phase are expected to be highly diverse, depending on the location of the intersection point, radiation hydrodynamics of the secondary shocks, viscous accretion of the nascent disk, as well as complex radiative transfer processes and viewing angles, see \citet{roth20_radiative_emission} for a review.

Another way of modeling the TDE system is to consider the evolution on timescales much longer than the time it takes for the orbits of the fallback material to circularize \citep{cannizzo90_disk_evolution, strubbe09_disk_wind, shen14_disk_evolution, balbus18_kerr_disk, mummery20_TDE_disk}. Under the assumption that the circularization is efficient \citep[as expected for sufficiently strong GR apsidal precession,][]{lu20_self_intersection, krolik20_TDE_variety}, the details of the disk formation process only play a minor role on the long-term evolution, provided that an order-unity fraction of the fallback material indeed joins the disk. This ``long-term averaging'' approach is also motivated by the late-time ($\gtrsim1\rm\, yr$) observations of TDEs: (1) optical transient surveys such as ZTF provide quasi-continuous (on a cadence of a few days) coverage of all nearby, bright TDEs for many years \citep{van_velzen20_ZTF_TDEs, hammerstein22_ZTF_phaseI}; (2) follow-up observations of many optically selected TDEs in the X-ray band show diverse properties (e.g., delayed X-ray emission, rapid variability, spectral evolution) \citep[][]{auchettl17_xray_properties, gezari17_as15oi_xray, margutti17_as15lh, wevers19_at2018fyk_xray_variable, wevers21_AT2018fyk, wen20_Xray_spectral_fitting, yao22_AT2021ehb};
(3) a few TDEs showed bright emission in the far-UV \citep{vanvelzen19_late_time_UV} and X-ray \citep{jonker20_late_time_xray} bands 5-10 years after the discovery; (4) eROSITA is currently scanning the whole sky at unprecedented depth at a $\sim$6-month cadence (with 8 epochs in total) and will hence provide the first large, X-ray selected TDE sample \citep{sazonov21_erosita_TDEs}. 

Many authors have also considered the possibility of a long-lived highly eccentric disk \citep{zanazzi20_eccentric_disk, linch21_eccentric_disk, liu21_eccentric_disk, chan22_eccentric_disk}, but this picture has not been realized in GR hydrodynamic simulations. It is expected that differential apsidal precession between fluid elements with different apocenter radii would lead to strong shocks which then drive circularization on a precessional timescale of $< t_{\rm fb}/\Phi_{\rm ap}$, which can be estimated to be $< 0.5\mr{\,yr}\, (M/10^6\,\msun)^{-1/6} (R_*/\rsun)^{5/2} (M_*/\msun)^{-4/3}$, where $t_{\rm fb} = \pi/\sqrt{2}\lrb{GM_*/R_*^3}^{-1/2}\lrb{M/M_*}^{1/2}$ is the orbital period of the most bound debris \citep{stone13_frozen-in}, $\Phi_{\rm ap}\simeq 3\pi \rg/\rp$ is the apsidal precession angle per orbit, $M$ is the BH mass, $M_*$ and $R_*$ are the stellar mass and radius, and we have taken the pericenter radius $\rp$ to be equal to the tidal disruption radius $\rT=R_*(M/M_*)^{1/3}$. It is possible that a minority of the bound gas acquire higher specific angular momenta than that of the initial star (as a result of angular momentum exchange between fluid elements) and hence their orbits could maintain moderately high eccentricities due to slower precession. It should also be noted that a high eccentricity might also be maintained by coherent precession in the entire disk if adjacent fluid elements are sufficiently strongly coupled by (an unusually high) viscosity. More detailed simulations are needed to verify these ideas. In the following, we focus on the case of a quasi-circular disk.

Our goal is to provide a concise review of the physics governing the long-term evolution of the disk formed in a TDE. The hope is to inspire more works along this interesting direction. The overall picture is that the disk evolution is governed by the viscous angular momentum transport and on-going fallback. The current model (based on a local $\alpha$-viscosity prescription) predicts a thermal-viscous instability for most TDEs with BH mass $\lesssim 10^7\msun$. This is similar to the widely believed disk-instability model for dwarf novae from cataclysmic variable systems \citep[see][for illustration of the ideas there]{lasota01_dwarf_novae, hameury20_dwarf_novae}, but a major difference in TDE disks is that it is the interplay between radiation and gas pressures that drives the instability.  Caveats on the assumptions in the current, highly simplified model will be pointed out for consideration by future works. After reviewing the model, we confront it with existing observations and then identify a number of major discrepancies that need to be reconciled in future works.

\section{2. Steady State of a Local Ring Region}\label{sec:steady_state_model}


Let us consider a small annulus of an axisymmetric disk located at radius $\rd$ from the BH of mass $M$, and the height-integrated surface mass density is denoted as $\Sigma$. The radius $\rd$ in consideration is far from the inner and outer boundaries. The length unit is the gravitational radius $\rg = GM/c^2$, where $G$ is the gravitational constant and $c$ is the speed of light. The goal is to find the steady-state solution for the local thermodynamic quantities (density $\rho$, pressure $P$, temperature $T$) in the disk mid-plane, under mass, momentum, and energy conservation laws \citep{frank02_accretion_disk_book}.

The vertical pressure scale height $H$ is given by the force balance between pressure gradient and gravity perpendicular to the disk mid-plane
\begin{equation}
  \label{eq:1}
  H = c_{\rm s}/\omgK = \sqrt{P/\rho}/\omgK,\ \ P = P_{\rm g} + P_{\rm rad}, \ \
  \omgK = \sqrt{GM/\rd^3}, 
\end{equation}
where $\omgK$ is the Keplerian frequency, $c_{\rm s}=\sqrt{P/\rho}$ is the isothermal sound speed, and the total pressure $P$ includes the contributions from gas pressure $P_{\rm g} = \rho k_{\rm B}T/(\mu \mp)$ and radiation pressure $P_{\rm rad} = aT^4/3$ ($a$ being the radiation density constant). We adopt a mean molecular mass of $\mu = 1.4/2.3$ for fully ionized gas with abundance ratio $n_{\rm He}/n_{\rm H}=0.1$. The gas-radiation mixture is assumed to be in local thermodynamic equilibrium (LTE) at the same temperature $T$, as a result of efficient gas-radiation coupling provided by absorption/emission and Comptonization. The gas density $\rho$ is related to the surface density by
\begin{equation}
  \label{eq:2}
  \Sigma = 2\rho H.
\end{equation}
It is useful to define a dimensionless scale height $\theta=H/\rd$, which can be obtained by combining Eqs. (\ref{eq:1}) and (\ref{eq:2}) into the following quadratic equation
\begin{equation}
  \label{eq:20}
  \theta^2 - b_1\theta - b_2 = 0 \Rightarrow \theta = {b_1 +
    \sqrt{b_1^2 + 4b_2} \over 2},\ \ b_1 = {2 aT^4 \rd^2\over 3GM\Sigma},
    \ \ b_2 = {k_{\rm B}T\rd\over GM\mu\mp}.
\end{equation}
We note that for fixed surface density and radius, the scale height is a monotonically increasing function of temperature and the density monotonically decreases with temperature.
Due to radial pressure gradient, the disk rotates at a sub-Keplerian frequency roughly given by
\begin{equation}
  \label{eq:Omega}
  \Omega \simeq {\omgK\over 1 + \theta^2}.
\end{equation}
The deviation from Keplerian rotation is included here since it does not add additional algorithmic complexity to the model.

The disk temperature determines the pressure scale height, which in turn controls the dynamical evolution of the system according to the $\alpha$-prescription for the viscosity \citep{shakura73_alpha_disk}
\begin{equation}
  \label{eq:4}
  \nu_{\rm vis} = \alpha H^2\omgK.
\end{equation}
and the corresponding viscous time is
\begin{equation}
  \label{eq:5}
  t_{\rm vis} = {\rd^2\over \nu_{\rm vis}} = {1\over \alpha \theta^2 \omgK}.
\end{equation}
The viscosity parameter is likely in the range $\alpha\in (\sim\!0.01, \sim\!0.1)$ \citep[see][for discussions of the hydrodynamic and magnetic components of the viscous stress]{sadowski16_deep_TDE}. We keep $\alpha$ fixed since the gas will always stay ionized in the parameter space we are interested in.
The accretion rate through radius $\rd$ is given by
\begin{equation}\label{eq:dotMacc}
    \dotMacc = |2\pi \rd v_{\rm r}\Sigma| = 3\pi \nuvis \Sigma,
\end{equation}
where $v_{\rm r}=-3\nu_{\rm vis}/(2\rd)$ is the radial inflow speed. A natural unit for the accretion rate is
\begin{equation}\label{eq:dotMEdd}
    \dotMEdd \equiv {10L_{\rm Edd}\over c^2} = 2.6\times10^{-2}M_6\,\msunyr, \ 
    L_{\rm Edd} = {4\pi GMc\over \kappa} =1.5\times10^{44}M_6\rm\, erg\,s^{-1},
\end{equation}
where $M_6=M/10^6\msun$, $L_{\rm Edd}$ is the Eddington luminosity and we adopt a typical accretion efficiency of $0.1$ in the above definition. Since we are interested in accretion rates near $\dotMEdd$, a natural unit for the surface mass density is
\begin{equation}\label{eq:Sigma_Edd}
    \Sigma_{\rm Edd}\equiv {\dotMEdd \over \alpha_{-1} \rg c} = {370\over \alpha_{-1}}\rm{\,g\,cm^{-2}},
\end{equation}
which roughly represents the surface density near $10\rg$ for an accretion rate near $\dotMEdd$, and we have used the notation $\alpha_{-1} = \alpha/0.1$.

An alternative ``$\beta$-viscosity'' prescription is that the viscous stress is only a fraction of the gas pressure instead of the total pressure  \citep{sakimoto81_beta_viscosity} --- under this prescription, the disk does not suffer from the thermal-viscous instability in the radiation-pressure-dominated regime \citep{lightman74_instability, shakura76_instability, piran78_instability} and remains stable throughout the evolution. The consequence of $\beta$-viscosity is that the disk evolves on a much longer timescales of decades \citep{cannizzo90_disk_evolution, vanvelzen19_late_time_UV}, and such a slowly evolving disk is disagreement with X-ray lightcurves of many eROSITA TDEs as well as optically selected ones. Radiation MHD simulations of shear flows, with explicit treatment of radiation, hydrodynamic and magnetic forces, demonstrated that the viscous stress is proportional to the total pressure as in the $\alpha$-disk model, and hence a radiation-dominated geometrically thin disk indeed undergoes thermal runaway evolution \citep{jiang13_thermal_instability}. Recent works by \cite{balbus18_kerr_disk, mummery20_TDE_disk} considered the evolution of a viscously spreading ring in the equatorial plane of a Kerr black hole, under the assumption that the viscous stress is a power-law function of radius without an explicit dependence on the local pressure --- their model may be applicable at very late time ($t\gg 10\rm\, yr$) in the self-similar phase \citep[as considered by][]{cannizzo90_disk_evolution} when the effects of continuous fallback are no longer important. Another possibility is that the disk can be stabilized by a strong magnetic field \citep{begelman07_magnetic_stable, sadowski16_magnetic_stable, jiang19_magnetic_stable, lancova19_magnetic_stable, mishra20_disk_magnetic_pressure}, which is possible due to the accumulation of magnetic flux near the event horizon throughout the accretion history. In fact, the relativistic jets in some TDEs \citep{bloom11_swiftj1644_57, burrows11_swiftj1644_57, cenko12_swiftj2058_05, brown15_swiftJ1112} are likely driven by the \citet{blandford77_BZ_jet} mechanism as a result of rapid BH spin and strong magnetic fields near the horizon \citep{tchekhovskoy14_mad_jet}. However, it is difficult for such a magnetic pressure-dominated state to be realized at radii far from the BH horizon. Based on these arguments, we proceed with the $\alpha$-viscosity. 


The disk temperature is set by energy conservation  --- the balance between heating rates and cooling rates per unit disk mass,
\begin{equation}
  \label{eq:3}
 T{\partial s\over \partial t} = Q^+ - Q^- =  (q_{\rm vis}^{+} + q_{\rm sh}^{+}) -
 (q_{\rm rad}^{-} + q_{\rm adv}^{-} + q_{\rm w}^{-}),
\end{equation}
where $s$ is the specific entropy.
We then note that the timescale for temperature variation, which is of the order $\omgK^{-1}$, is much shorter than the variational timescale for the heating and cooling rates (viscous time), so the disk material quickly reaches thermal equilibrium with $\partial s/\partial t = 0$. Our model is not to be trusted on a timescale of $\omgK^{-1}$ anyway (due to ignorance of the disk formation physics), so we only consider the quasi-thermal equilibrium solution as given by $Q^+ - Q^- = 0$.

Below, we estimate each of the terms in the heating and cooling rates.
The viscous heating rate is
\begin{equation}
  \label{eq:6}
  q_{\rm vis}^+ = {9\over 4}\nu_{\rm vis}\Omega^2 = {9\over 4} {GM/\rd
  \over (1 + \theta^2)^2 t_{\rm vis}}.
\end{equation}
The shock heating is due to circularization of newly infalling
material
\begin{equation}
  \label{eq:8}
  q_{\rm sh}^{+} = f_{\rm sh} {GM/\rd\over 2} {\dot{\Sigma}_{\rm fb}\over
    \Sigma},
\end{equation}
where $\dot{\Sigma}_{\rm fb}$ is the mass fallback rate per unit disk area, and $f_{\rm sh}< 1$ describes the fraction of the energy dissipated by the circularization shocks that is used to heat up the disk gas. Here, ``circularization shocks'' include all dissipation mechanisms that convert the highly eccentric ($1-e\ll 1$) orbits of the infalling material into quasi-circular ones ($e\lesssim 0.5$). As demonstrated in \citet{bonnerot21_disk_formation}, in the case of a strong stream-self-crossing shock that diverts the fallback material in a quasi-spherical manner, most of the orbital energy is dissipated by secondary shocks that are located at roughly a few scale-heights away from the disk mid-plane. Nearly all the shock-generated heat is quickly radiated away in the vertical direction instead of being deposited into the existing disk material, indicating $\fsh\ll 1$. At the current moment, since the hydrodynamics of the disk formation process are not well understood, we do not have a good prescription for $f_{\rm sh}$. Future works on the interactions between the fallback gas and the existing disk are needed to provide a more realistic picture. It should be noted that, in a different context, the interaction between the fallback stream and a pre-existing disk (in an active galactic nucleus) has been studied by \cite{kelley14_TDE_AGN, chan21_TDE_inAGN}, but the setup there is a very dense and thin stream penetrating through a disk at relatively low accretion rates --- very different from the situation of disk formation in TDEs.


The radiative cooling rate is
\begin{equation}
  \label{eq:9}
  q_{\rm rad}^{-} = {U_{\rm rad} c\over \rho H (\tau + 1)},\ \ \tau =
  \rho \kappa H, \ \ U_{\rm rad} = aT^4,
\end{equation}
where we have included photon diffusion in the optically thick regime for $\tau\gg 1$ and direct escaping in the optically thin regime for $\tau \ll 1$ (in fact we are exclusively in the $\tau\gg 1$ regime in the first decade after a TDE). The Rosseland-mean opacity $\kappa(\rho, T)$ is given by the OPAL table \citep{iglesias96_OPAL} at solar metallicity. For the majority of the relevant parameter space, the opacity is dominated by Thomson scattering $\kappa \simeq 0.34\rm\, g\, cm^{-2}$. In super-Eddington disks, most of the dissipation does not occur right at the disk mid-plane and the radiative flux in the vertical direction has comparable contributions from advection (due to convective motion) and diffusion \citep{jiang13_thermal_instability}. We expect that these effects will lead to faster radiative cooling than in eq. (\ref{eq:9}). This is another caveat to keep in mind in future works.

The radial inflow advects heat to smaller radii \citep{narayan94_adaf1} and hence contributes to an advective cooling rate approximately given by
\begin{equation}
\label{eq:qadv}
  q_{\rm adv}^{-} = v_{\rm r}T{\partial s\over \partial r} \simeq
  {3\over 2} \lrb{{U\over P} - 1} \theta^2 {GM/\rd \over t_{\rm vis}},
\end{equation}
where $s$ is the specific entropy and $v_{\rm r} = -3\nu_{\rm vis}/(2\rd)$ is the radial velocity of the mass inflow. To obtain the second expression in eq. (\ref{eq:qadv}), we have made use of $T\d s = \d h - \d P/\rho$, $h=(U+P)/\rho\propto c_{\rm s} \propto r^{-1}$ being the specific enthalpy ($U$ being the thermal energy density) and $\partial h/\partial r\simeq -h/\rd$, and we have taken an approximate radial pressure scaling of $P\propto r^{-2}$ (and hence $\partial P/\partial r\simeq -2P/\rd$) in between extreme cases of no wind loss ($P\propto r^{-2.5}$) and strongest wind loss ($P\propto r^{-1.5}$) \citep{blandford99_ADIOS}.

Finally, we include the effects of possible disk wind which carries energy away from the system and gives a cooling rate
\begin{equation}
  \label{eq:qw}
  q_{\rm w}^{-} = \fw {GM/\rd\over \tvis},
\end{equation}
$f_{\rm w}$ is a strength parameter that is roughly the fraction of the disk mass evolution rate that is carried away by wind $\dotMw\simeq \fw \Md/\tvis$ (provided that the asymptotic energy of the wind is $GM/\rd$). In the limit of a geometrically thin disk $\theta\ll 1$, the disk material is strongly bound and hence we expect wind cooling to be unimportant $(\fw\ll 1)$ as compared to radiative cooling. In the limit of a thick disk $\theta\sim 1$, it is possible that the wind cooling rate is much higher than the radiative cooling rate.
The detailed (radiation$+$GRMHD) wind launching physics in super-Eddington accretion disks are still uncertain, due to a lack of large-scale and long-term simulations \citep{jiang19_superEdd_disk}. A physically motivated prescription in the literature is the ``Bernoulli-limited wind'' \citep{margalit16_WD_NS_TDE}
\begin{equation}
  \label{eq:12}
  \fw = \mr{Sigmoid}(\Be),
\end{equation}
where the Sigmoid function approaches 1 when the argument is positive and 0 when the argument is negative, and the dimensionless Bernoulli number is defined as the total energy (sum of enthalpy and kinetic/potential energies) normalized by the gravitational potential energy
\begin{equation}
  \label{eq:13}
  B_{\rm e} = {(U+P)/\rho + \Omega^2\rd^2/2 - GM/\rd \over GM/\rd} \approx 
  {U\over P} \theta^2 - 0.5,
\end{equation}
where $U=3\rho\kB T/(2\mu\mp) + aT^4$ is the thermal energy density [in the second expression above we have ignored high-order terms $\mc{O}(\theta^4)$]. The Bernoulli number becomes positive when $q_{\rm adv}^-\gtrsim q_{\rm rad}^-$ in the Advection-Dominated Accretion Flow (ADAF) regime \citep{narayan94_adaf1, narayan95_adaf2}. A positive Bernoulli number drives the gas unbound and we expect a large fraction of the disk mass to be lost on a viscous time, as motivated by numerical simulations of radiatively inefficient disks \citep[e.g.,][]{stone99_ADIOS, yuan12_outflow, narayan12_ADAF_simulation} \citep[see][for a review]{yuan14_review}. In this case, the wind effectively cools the disk to a negative Bernoulli number. Practically, we take the Sigmoid function to be $\mr{Sigmoid}(\Be) = [1 + \mr{e}^{-\Be/\Delta\Be}]^{-1}$ such that the transition occurs in a narrow range of $\Delta B_{\rm e} = 0.1$ (the results depend little on this choice).

With all the expressions for the heating and cooling rates,
we define a dimensionless function
\begin{equation}
  \label{eq:18}
  \begin{split}
  g(T; \Sigma, \rd) &\equiv {(Q^+-Q^-)\omgK^{-1}\over GM/\rd} \\
  &= {f_{\rm
      sh}\dot{\Sigma}_{\rm fb}\over 2\Sigma \omgK}  - {2 U_{\rm rad}c \over \Sigma \rd^2 \omgK^3 (\tau + 1)}
   +{\alpha\theta^2}\left[{9\over 4(1+\theta^2)^2} - {3\over 2}\lrb{{U\over P}-1}\theta^2 - \fw\right],
  \end{split}
\end{equation}
and the solutions of thermal equilibrium are obtained by solving $g(T) = 0$. Since $g(T)$ (for fixed $\Sigma$ and $\rd$) is not a monotonic function, it is possible to have multiple solutions.
An unstable solution has the property that if we perturb the temperature $T$ slightly, the perturbation will grow exponentially.
The condition for thermal stability is
\begin{equation}
  \label{eq:14}
  \left.{\partial g\over \partial T}\right|_{\Sigma, \rd} < 0,
\end{equation}
which means that, for fixed disk surface density and radius, if the temperature is slightly increased (or decreased), the system will cool (heat) back to the equilibrium state of zero net heating rate $Q^+-Q^-=0$.

To illustrate the simple physics of thermal instability, let us consider the situation of a radiation-dominated disk with $U\approx U_{\rm rad}$ and $P \approx U_{\rm  rad}/3$, ignoring mass fallback and wind cooling for now (by taking $f_{\rm sh}=\fw=0$). In this case, we have $U_{\rm rad} = 3GM\rho \theta^2/\rd$, which means
\begin{equation}
  \label{eq:19}
  g = {3\alpha\theta^2}\left({3\over 4(1+\theta^2)^2} - \theta^2\right) -
  b\,\theta, \ \ b = {3c \over (\tau+1)\rd\omgK}.
\end{equation}
It is clear that a thin disk ($\theta\ll 1$) in this case is thermally unstable, because for fixed disk mass and radius (and hence $b$ is fixed), the solution to $g=0$ is $\theta \approx 4b/9\alpha$, and it is easy to show $(\partial g/\partial \theta)|_{\theta=4b/9\alpha} \approx b > 0$ which means $\partial g/\partial T>0$ --- meaning that an isolated thin radiation-pressure-dominated disk is unstable.



\begin{figure}
\centering
\includegraphics[width=0.49\textwidth]{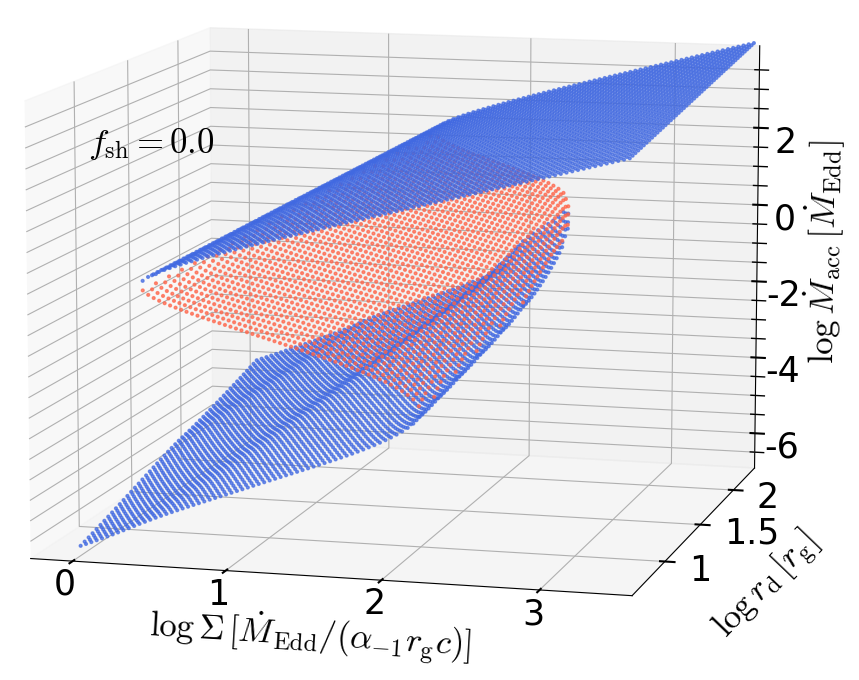}
\includegraphics[width=0.49\textwidth]{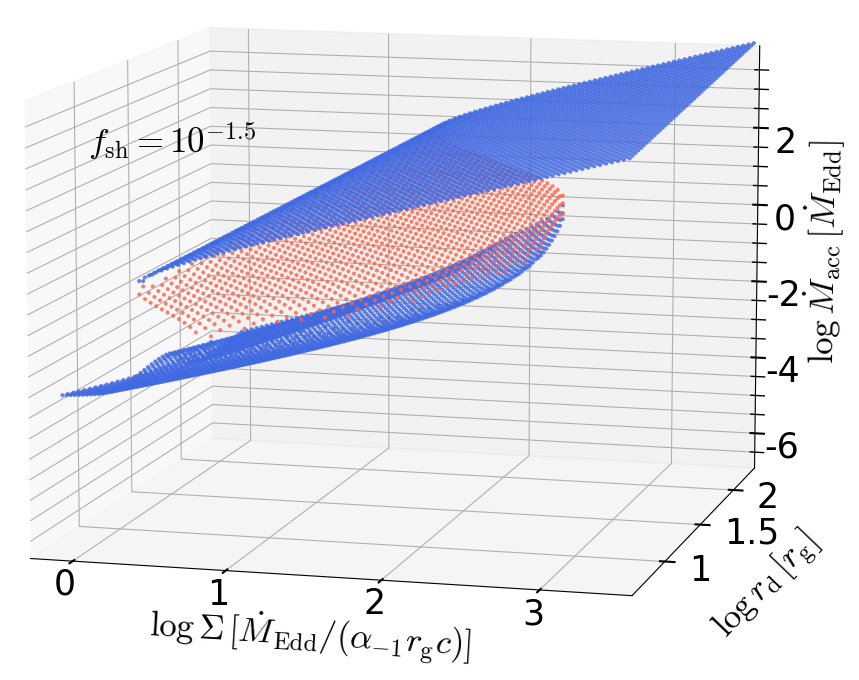}
\caption{Thermal equilibrium solutions ($g=Q^+ - Q^-=0$) at different radii $\rd$ in units of gravitational radius $\rg$ and surface densities $\Sigma$ in units of $\Sigma_{\rm Edd}=\dotMEdd/(\alpha_{-1}\rg c)$ (as defined in eq. \ref{eq:Sigma_Edd}). The mass accretion rate $\dotMacc$ (eq. \ref{eq:dotMacc}) is expressed in units of $\dotMEdd=10L_{\rm Edd}/c^2$. The solutions shown in red are thermally unstable such that the disk will undergo a state transition in the vertical direction (preserving $\Sigma$ and $\rd$) onto one of the blue surfaces of stable solutions. In the left panel, the effects of shock heating is ignored ($\fsh=0$), whereas in the right panel, a small amount of shock heating is considered ($\fsh = 10^{-1.5}$) assuming a fallback rate near the Eddington level $\pi \rd^2\dot{\Sigma}_{\rm fb}=\dotMEdd$. A finite shock heating rate increases the accretion rate in the lower (gas-pressure-dominated) stable branch, whereas the upper stable branch is nearly unaffected. In both panels, the BH mass is $M = 10^6\msun$ and the viscous parameter is taken to be $\alpha=0.1$. The results, expressed in the above dimensionless units, depend very weakly on $M$ and $\alpha$.
}
\label{fig:disksolution}
\end{figure}

The solutions to $g(T) = 0$ for a broad range of radii and surface densities are shown in Fig. \ref{fig:disksolution}. The left panel of this figure is similar to Fig. 3 of \citet{shen14_disk_evolution}, who did not consider the effects of shock heating (by taking $\fsh=0$). The right panel shows the solutions for a modest amount of shock heating rate ($\fsh=10^{-1.5}$ with a local fallback rate near the Eddington level). We highlight the stable ($\partial g/\partial T <0$) solutions in blue and unstable ($\partial g/\partial T <0$) ones in red. When in a steady state, different regions of the TDE disk are expected to lie on the blue surfaces --- it is possible that parts of the disk in a certain radius range are on the lower branch whereas other parts are on the upper branch. 

\begin{figure}
\centering
\includegraphics[width=0.95\textwidth]{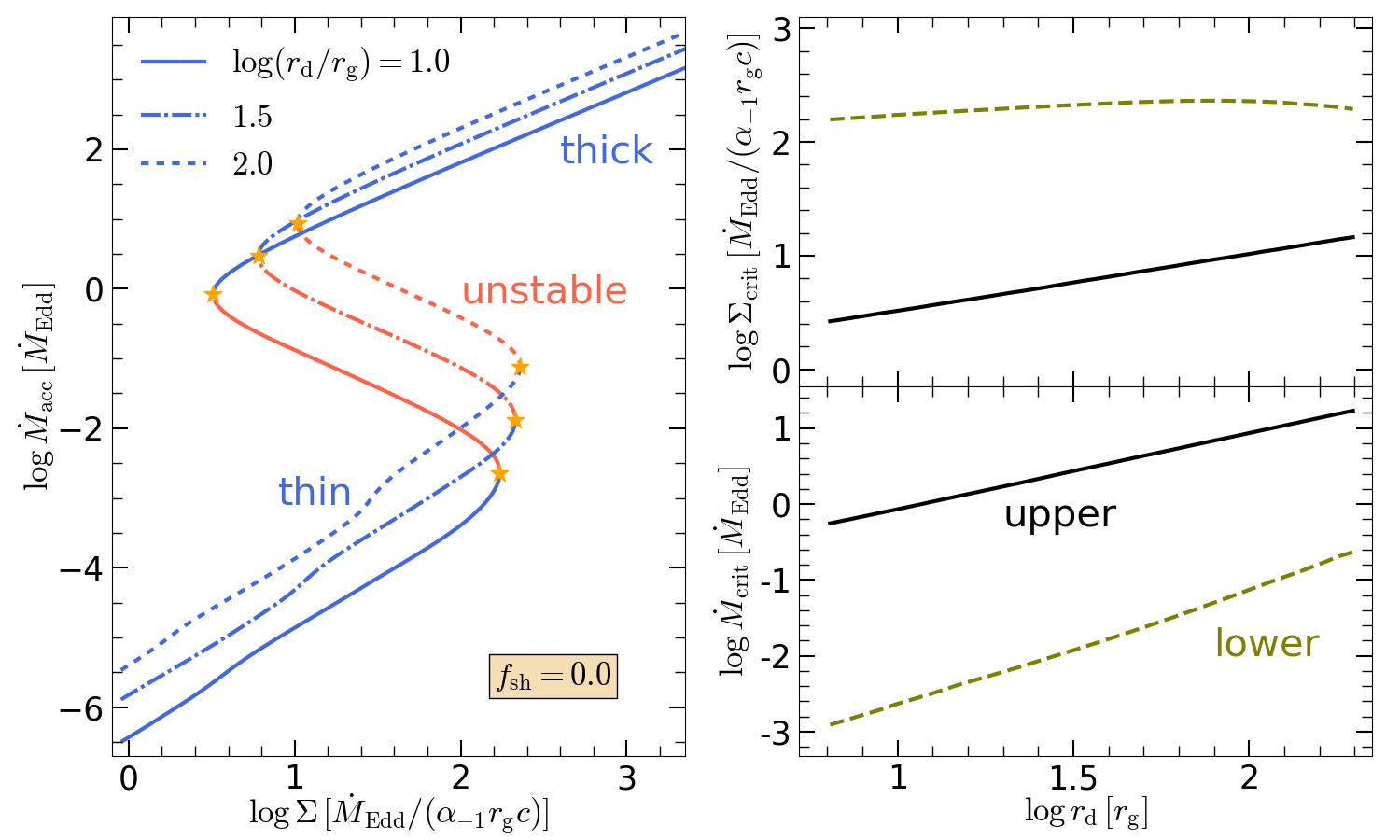}
\caption{Left panel: Thermal equilibrium solutions ($g=0$) at three different disk radii $\rd=10, 30, 100\rg$. The solutions shown by the red (or blue) lines are unstable (stable). The orange stars mark the boundaries between the stable and unstable solutions where $\lrb{\partial g/\partial T}_{\Sigma, \rd} = 0$. Bottom right panel: The accretion rates at the two instability boundaries --- the upper one is adjacent to the geometrically thick branch and the lower one connects to the geometrically thin branch. Upper right panel: The surface densities at the two instability boundaries. In this figure, we ignore shock heating by taking $\fsh=0$, and adopt BH mass $M=10^{6}\msun$ and viscous parameter $\alpha=0.1$ (the dimensionless results depend weakly on these two parameters).
}
\label{fig:boundary}
\end{figure}

\begin{figure}
\centering
\includegraphics[width=0.95\textwidth]{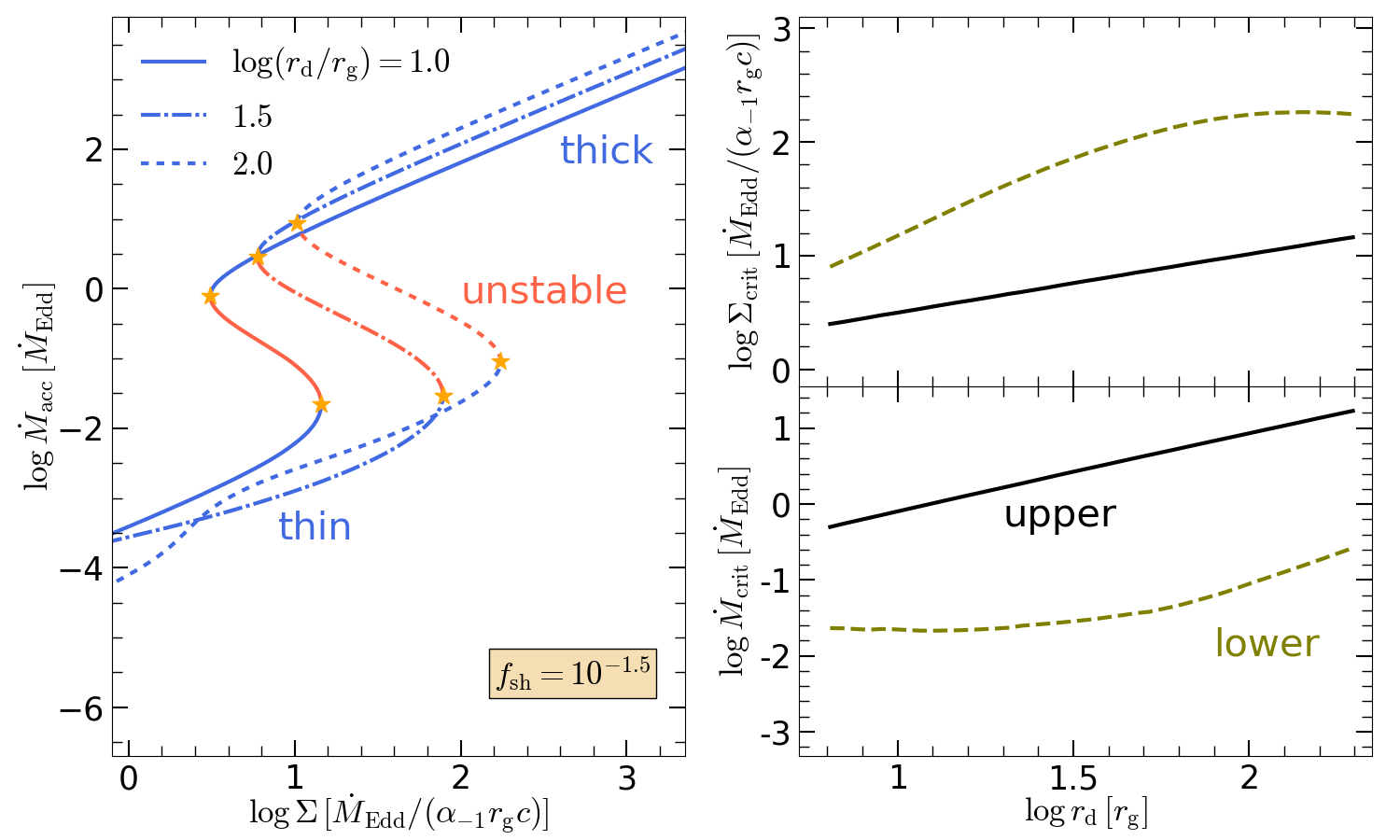}
\caption{The same as Fig. \ref{fig:boundary} but with a finite shock heating rate with $\fsh=10^{-1.5}$ and a fallback rate near the Eddington level $\pi \rd^2\dot{\Sigma}_{\rm fb} = \dotMEdd$.
}
\label{fig:boundary_fsh}
\end{figure}

We also show the solution to $g(T) = 0$ for a number of fixed radii $\log(\rd/\rg) = 1, 1.5, 2$ in the upper panels of Fig. \ref{fig:boundary} (for $\fsh=0$) and Fig. \ref{fig:boundary_fsh} (for $\fsh\neq0$). When the gas at a given radius evolves to approach the red-blue boundary, the local disk undergoes a state transition with a sudden jump in accretion rate while preserving both $\Sigma$ and $\rd$. The basic expectation in a TDE is that, on the timescale of a few years, the fallback rate gradually drops from super-Eddington to sub-Eddington values, and the disk may undergo state transitions as in the case of Galactic BH X-ray binaries \citep{fender04_xrb, remillard06_xrb}. One possible manifestation of such a transition is the steep X-ray flux decline (by a factor of $10^2$) in the two jetted TDEs Swift J1644+57 \citep{zauderer13_1644jet_shut_off} and Swift 2058+05 \citep{pasham15_2058jet_shut_off} at a few hundred days (in the host galaxy rest frame) following the more gradual X-ray flux decline. Although the jet launching processes and X-ray radiation mechanisms are still debated \citep[see][for a review]{decolle20_tde_jets}, the sudden drop in X-ray flux is most naturally explained as the jet shutting off when the innermost regions of the accretion disk transitions from a hot, geometrically thick state to a cold, thin one \citep{shen14_disk_evolution, tchekhovskoy14_mad_jet}. In the next section, we carefully consider this possibility in the context of long-term disk evolution in TDEs.


The boundaries of stability are located at two lines given by the joint solutions to $\lrb{\partial g/\partial T}_{\Sigma, \rd} = 0$ and $g=0$. The upper boundary connects the unstable branch (shown in red) to the high accretion rate solutions that are geometrically thick ($\theta\sim 1$) and radiation-pressure dominated. The lower boundary connects the unstable branch to the low accretion rate solutions that are geometrically thin ($\theta\ll 1$) and gas-pressure dominated. The critical accretion rates and critical surface densities at these two boundaries are shown in the right panels of Fig. \ref{fig:boundary}. When shock heating is ignored, our numerical results for the critical values can be described by the following approximations (for $\fsh=0$)
\begin{equation}\label{eq:Mdotcrit}
\begin{split}
    \dot{M}_{\rm crit}^+(\mbox{upper}) & \approx \dotMEdd \,(\rd/10\rg), \
    \dot{M}_{\rm crit}^-(\mbox{lower}) \approx 2\times10^{-3}\,\dotMEdd \, (\rd/10\rg)^{3/2},\\
    \Sigma_{\rm crit}^+ (\mbox{upper}) &\approx 3.0\, \Sigma_{\rm Edd} (\rd/10\rg)^{1/2},\ 
    \Sigma_{\rm crit}^- (\mbox{lower}) \approx
    1.6\times10^2\,\Sigma_{\rm Edd} (\rd/10\rg)^{1/5}\alpha_{-1}^{1/7}M_6^{1/8},
\end{split}
\end{equation}
where the units $\dotMEdd$ and $\Sigma_{\rm Edd}$ are defined in eqs. (\ref{eq:dotMEdd}) and (\ref{eq:Sigma_Edd}). For a finite shock heating rate, only the critical values at the lower boundary ($\dot{M}_{\rm crit}^-$ and $\Sigma_{\rm crit}^-$) are affected (see the right panels of Fig. \ref{fig:boundary_fsh}).



In the next section, we consider a simple model for the dynamical evolution of the TDE disk taking into account on-going mass fallback and viscous accretion.

\section{3. Dynamical Evolution}\label{sec:evolution}

\subsection{3.1 Piecewise Steady-state One-zone Model}\label{sec:outer_disk_model}

In the previous section, we have obtained the thermodynamic equilibrium solution at a single radius for a given surface mass density $\Sigma$. In the realistic situation, one would hope to solve the time evolution of the surface mass density at all radii $\Sigma(r, t)$. The simplest case is to consider a ``one-zone'' model for a disk where most of the disk mass $\md$ is contained near the outer disk radius $\rd$, where the surface mass density is given by
\begin{equation}
    \Sigma(\rd) \simeq {\md \over \pi \rd^2}.
\end{equation}
After we have obtained the mass accretion rate near the outer disk at a given time, then this can be used as the boundary condition to solve for a steady-state disk profile in the inner regions at $r\ll \rd$ where the relevant timescales are much shorter than the evolutionary time of the outer disk. This approach was taken by \cite{strubbe09_disk_wind, shen14_disk_evolution}, and earlier by \cite{cannizzo90_disk_evolution} in a simpler set-up without continuous gas supply. As we see later (\S 3.2), this ``one-zone'' model is highly incomplete when confronted with observations. We note that this crude model has been used in other contexts \citep{kumar08_GRB_disk, metzger08_BNS_disk} where it does capture most of the physics relevant for the long-term evolution of those systems.


The mass and radius of the outer disk evolve on the viscous time scale $t_{\rm vis}(\rd)$ (eq. \ref{eq:5}), according to mass and angular momentum  conservations. The disk mass evolves due to the fallback of the bound debris and viscous accretion
\begin{equation}
  \label{eq:15}
  \dotMd = \dotMfb - \dotMacc, \ \dotMacc \simeq {\md/t_{\rm vis}},
\end{equation}
where a fraction of $\dotMacc$ could be due to wind loss when the Bernoulli number is positive. Here the accretion rate $\dotMacc$ is only a rough estimate, because the radial velocity in the outer disk bifurcates in a way that some gas moves inwards ($v_{\rm r}<0$) and the rest moves outwards ($v_{\rm r}>0$). The fallback rate is given by the properties of the disrupted star (mass $M_\star$ and radius $R_{\star}$) as well as the BH mass. For most TDEs, the BH spin only weakly affects the fallback rate \citep{gafton19_kerr_disruption}. Later on, we take the fallback rate from the Newtonian hydrodynamic simulations by \citet{law-smith20_STAR_library}. The numerical fallback rate is similar to that based on the ``frozen-in'' approximation \citep{stone13_frozen-in} at late time,
\begin{equation}
  \label{eq:fallback_rate}
 \dotMfb \sim  {M_\star\over 3t_{\rm fb}} (t/t_{\rm
  fb})^{-5/3}, \mbox{ for $t\gtrsim t_{\rm fb}$ and } t_{\rm fb} \simeq 41\mr{\,d}\, \left(M\over10^6\msun\right)^{1/2}
\left(M_\star\over\msun\right)^{-1} \left(R_\star\over R_\odot\right)^{3/2}.
\end{equation}
It should be noted that the underlying assumptions here are: (i) a large fraction of the fallback material joins the disk instead of being driven away as an outflow or falling into the event horizon before orbital circularization, and (ii) the mass feeding rate to the disk tracks the fallback rate.

Regarding assumption (i), it has been demonstrated, by \cite{jiang16_self-intersection} using numerical simulations and by \cite{lu20_self_intersection} semi-analytically, that a fraction of the fallback material can become unbound due to a strong self-crossing shock. This is because the gas is only marginally bound before entering the self-crossing shock and can easily gain enough energy from pressure forces in the post-shock region to be ejected to infinity \citep[Fig. 8 of][shows the effects of the energy redistribution by a strong self-crossing shock]{lu20_self_intersection}. However, since the unbound fraction cannot exceed 50\%, we ignore this factor since there are larger uncertainties elsewhere in the current consideration. The angular momentum redistribution, however, can cause significant changes in the radius where the material circularizes, and this effect will be considered later. It should also be noted that, when the self-crossing shock occurs near the event horizon, a significant fraction of the shocked gas can directly plunge into the event horizon, but since these cases are relatively rare, we do not discuss them here (the accretion disk likely does not play the dominant role in these TDEs, and their luminosity is likely dominated by shock energy dissipation). 

Regarding assumption (ii), it is well known that Lense-Thirring precession by a rapidly spinning BH can cause the stream to miss self-crossing for many orbits \citep{dai13_Kerr_orbits, guillochon15_LT_precession, batra21_GR_precession}, and when this occurs, self-crossing is delayed and the feeding rate will be reduced. The delay time depends on the interplay between GR precesssions and the hydrodynamic evolution of the thickness of the thin stream. Without a better model at this moment, we proceed with the simple fallback rate and hence the results are only applicable to the cases with either slowly spinning BHs or low inclination angles between the initial star's angular momentum and the BH's spin. 

The disk angular momentum is given by $J_{\rm d} = \md \Omega\rd^2$, where $\Omega$ is given by eq. (\ref{eq:Omega}). We ignore the small amount of angular momentum loss due to accretion through the event horizon, and further assume that the wind (with a mass loss rate $\dotMw\simeq \fw \dotMacc$) carries the same specific angular momentum as the disk material, then the time evolution is given by
\begin{equation}
  \label{eq:17}
  \dot{J}_{\rm d} = \dot{J}_{\rm fb}-\dot{M}_{\rm w} \Omega \rd^2,
\end{equation}
where $\dot{J}_{\rm fb} = \sqrt{GM\rc} \dotMfb$ is the angular momentum newly added by the fallback material and $\rc$ is the ``circularization radius'' --- the radius of a circular Keplerian orbit with the same specific angular momentum as the fallback material.

By ignoring a small $\d\theta^2/\d t$ term, we obtain $\dot{J}_{\rm d}/J_{\rm d} = \dot{r}_{\rm d}/2\rd + \dotMd/\md$. Combining these expressions, we obtain the radial expansion rate of the outer disk
\begin{equation}
  \label{eq:17}
  \dot{r}_{\rm d} = 2\rd\left[{1-\fw\over t_{\rm vis}} + {\dotMfb\over
      \md} \left({(1 + \theta^2)\sqrt{\rc\over \rd}} - 1\right)\right].
\end{equation}
Since the disk can adjust its temperature in a dynamical time $\omgK^{-1}$ ($=$the sound crossing time in the vertical direction), which is much shorter than the viscous time $t_{\rm vis}$, we seek for the thermodynamic equilibrium solution by solving $g(T;\md, \rd)=0$ for the temperature at each time $t$ with known $\md(t)$ and $\rd(t)$, and then integrate eqs. (\ref{eq:15}) and (\ref{eq:17}) forward in time to obtain the $\md(t+\Delta t)$ and $\rd(t+\Delta t)$ at the next timestep, and then the temperature can be solved again. The initial conditions are taken to be $\md(t=0) = M_{\rm d,0} =3\times 10^{-3}\msun$ and $\rd(t=0) = \rc$. The choice of $M_{\rm d,0}$ is ``forgotten'' by the system in the first few days and hence does not affect the late-time behavior of the disk.

\subsection{3.2 Results and Comparison with Observations}\label{sec:results}

\begin{figure}
\centering
\includegraphics[width=0.49\textwidth]{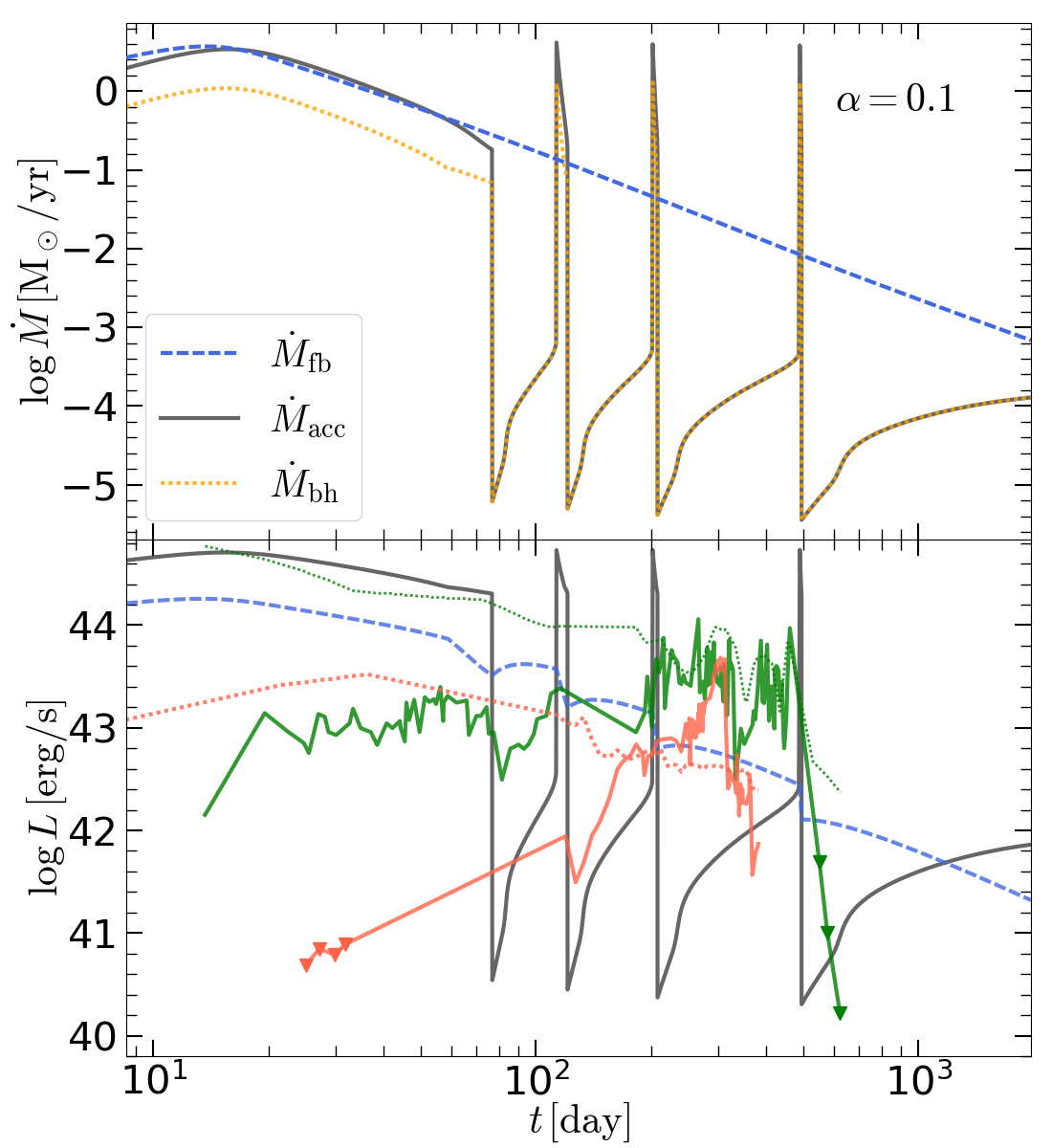}
\includegraphics[width=0.49\textwidth]{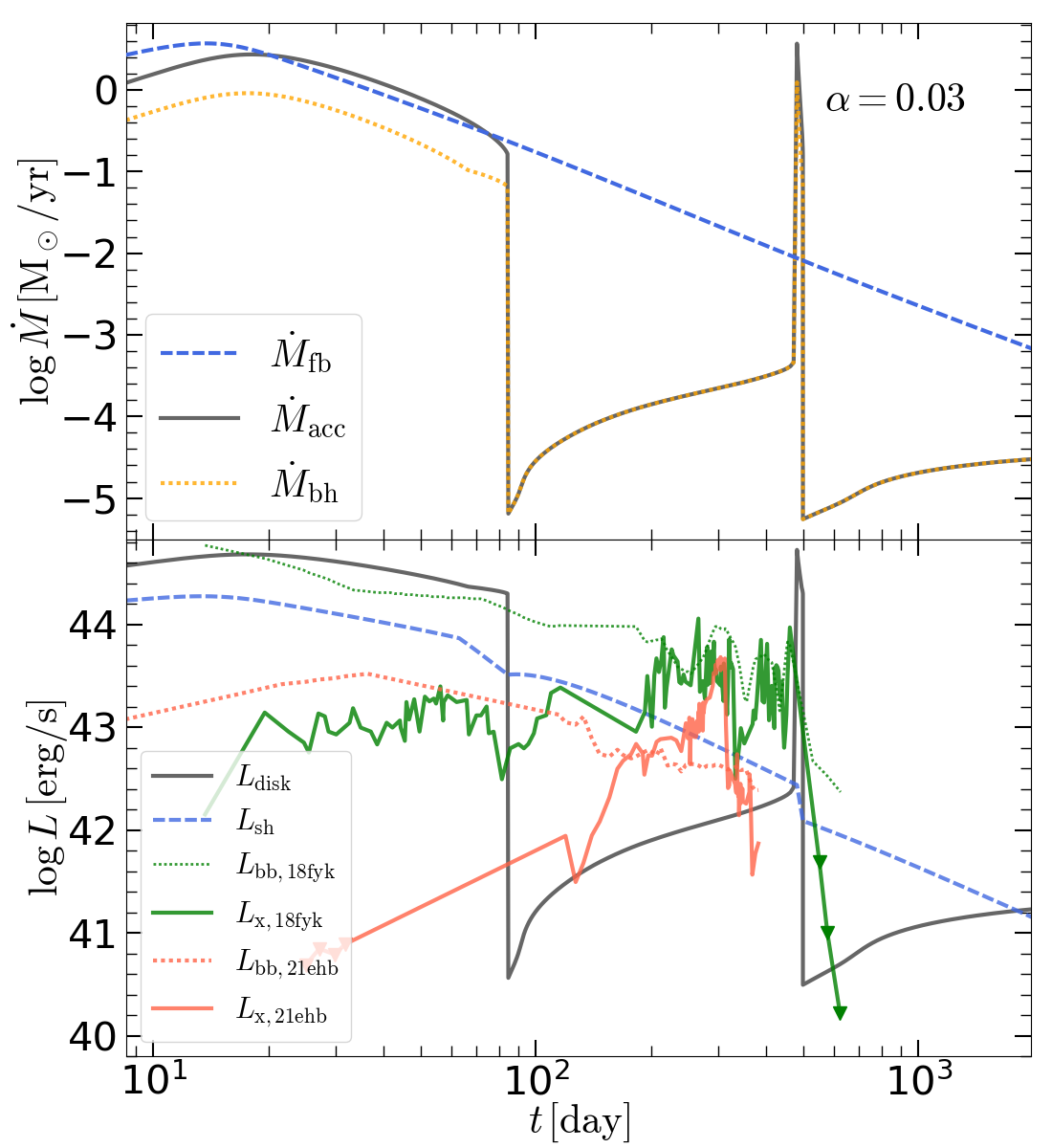}
\caption{Disk evolution for the disruption of a Sun-like star by a BH of $M=10^6\msun$. The horizontal axis shows the time since the fallback of the most bound material. The circularization radius of the fallback debris is taken to be $\rc = \rT$, where $\rT = (M/M_*)^{1/3}\rsun$ is the tidal disruption radius. The viscosity parameter is taken to be $\alpha=0.1$ and $0.03$ in the left and right panels, respectively. The effects of shock heating of the existing disk gas by the fallback material are ignored (by taking $\fsh=0$). The upper panels show the accretion rate of the outer disk $\dotMacc = \md/t_{\rm vis}$ (black solid lines), the BH's accretion rate $\dot{M}_{\rm bh}$ (orange dotted), and the fallback rate $\dotMfb$ (blue dashed lines). The lower panels show the emerging luminosities from the \textit{entire} accretion disk (black solid lines) and from the circularization shocks $L_{\rm sh}$ (blue dashed lines). We also show the observations of two TDEs, AT2018fyk \citep[green lines,][]{wevers21_AT2018fyk} and AT2021ehb \citep[red lines][]{yao22_AT2021ehb}, whose UV/optical ``blackbody'' luminosities $L_{\rm bb}$ are show by dotted lines and X-ray luminosities $L_{\rm x}$ are shown by solid lines (they are the same in the left and right panels). We omit error bars for clarity, and useful upper limits are shown by downwards triangles. The observational time is since discovery (in the host galaxy comoving frame). The X-ray lightcurves show steep declines at $t\simeq 320\rm\,d$ (both 18fyk and 21ehb), $t\simeq 370\rm\,d$ (21ehb), and $t\sim 500\rm\,d$ (18fyk), indicating state transitions in the accretion disks.
}
\label{fig:evolution}
\end{figure}

In Fig. \ref{fig:evolution}, we show the disk evolution based on the model described in \S 3.1 for a typical TDE of a Sun-like star ($M_*=1\msun$ at a main-sequence age of 5 Gyr) as computed by $\mathtt{MESA}$ \citep{paxton19_mesa} disrupted by a $10^6\msun$ BH. The fallback rate is taken from \cite{law-smith20_STAR_library}. The initial stellar pericenter distance is $\rp = \rT/2$ (for a penetration parameter of $\beta\equiv \rT/\rp=2$) such that the star is fully disrupted and that the circularization radius is $\rc=\rT$. We show the results for two different choices of $\alpha=0.1$ (left panel) and $0.03$ (right panel). In both panels, we ignore shock heating by taking $\fsh=0$ due to the lack of a self-consistent model for it at this moment, and we will discuss the effects of shock heating later in this subsection.


Once the outer disk evolution [$\Md(t)$ and $\rd(t)$] is obtained, we infer the accretion rate onto the BH using the following arguments. Suppose the outer disk radius is $\rd$ and accretion rate is $\dotMacc$ at a given time, the system will have a spherization radius given by \citep{shakura73_alpha_disk}
\begin{equation}
    \rsph = \min\lrb{\rd, {\dotMacc \kappa \over 4\pi c}},
\end{equation}
The meaning of the spherization radius is that the disk becomes geometrically thick at radii $r<\rsph$ and a strong wind is produced according to our Bernoulli criterion. Accounting for the wind loss, the accretion rate onto the BH may be estimated by \citep{blandford99_ADIOS}
\begin{equation}
    \dotMbh \simeq \dotMacc \lrb{10\rg\over \rsph}^s,
\end{equation}
where the numerator inside the power-law scaling is taken to be slightly larger than the radius of the innermost stable circular orbit (ISCO) and the power-law index $s$ is likely between 0.3 and 1 as shown by numerical simulations \citep{yuan12_outflow}. Since TDE disks are generally compact, with $\rd\lesssim 100\rg$, the uncertainty in $s$ does not affect our discussion qualitatively and we choose $s=0.5$ as a representative value. The total luminosity emerging from the accretion disk can then be estimated by
\begin{equation}
    L_{\rm disk}\simeq 0.1\dotMbh c^2 \min\lrsb{1, (\rtrin/10\rg)^{-2/3}},\  \rtrin = {\dotMbh \kappa\over 4\pi c},
\end{equation}
where we have included possible adiabatic loss below the photon trapping radius $\rtrin$ for the outflowing material from the inner disk near the ISCO. We note that a rapidly spinning BH will likely launch a pair of relativistic jets whose luminosity may be higher than $L_{\rm disk}$ taken here, provided that there is a high magnetic flux threading the event horizon \citep[e.g.,][]{tchekhovskoy11_MAD_efficiency, tchekhovskoy14_mad_jet, narayan22_MAD_efficiency}.

The freshly infalling material joins the existing disk as a result of shock interactions, which convert the highly eccentric ($1-e\ll 1$) orbits into quasi-circular ones ($e\lesssim 0.5$).
The gas heated by the circularization shocks produces radiation which escapes the system by diffusion \citep{piran15_disk_formation, bonnerot21_disk_formation}, and the emerging luminosity can be estimated by
\begin{equation}
    L_{\rm sh} = \dotMfb {GM\over 2\rd} \min\lrsb{1, (\rtrfb/\rd)^{-2/3}}, \ \rtrfb = {\dotMfb \kappa\over 4\pi c},
\end{equation}
we have included potential adiabatic losses below the photon trapping radius of the fallback material.


We see many interesting features in Fig. \ref{fig:evolution}, most of which have been identified by the previous works \citep{strubbe09_disk_wind, shen14_disk_evolution}. In the following, we compare these features with recent observations in the past few years and identify dis/agreements. 

(1) At early time $t\lesssim 100\rm\, d$, the outer disk is geometrically thick and has high accretion rates comparable to the fallback rate. During this phase, most of the fallback mass is lost from the system as disk wind. The radiation from the accretion disk and circularization shocks suffer strong adiabatic losses due to photon trapping in the disk wind and the fallback material --- the emerging bolometric luminosity is mildly super-Eddington.
This is in qualitative agreement with observations of some of the brightest TDEs \citep{van_velzen20_ZTF_TDEs, hammerstein22_ZTF_phaseI}, although the early-time bolometric luminosity of most optically-selected TDEs is dominated by UV/optical emission whereas their X-ray fluxes are subdominant. To reproduce the spectrum in the optical/near-UV bands, the photons emerging from the accretion disk or circularization shocks need to be reprocessed (i.e., absorbed and re-emitted) by an optically thick gas layer at large distances of $10^{14}$--$10^{15}\rm\, cm$ \citep{loeb97_Eddington_envelope, metzger16_reprocessing, roth16_reprocessing}. It has been proposed \citep{lu20_self_intersection} that the reprocessing layer originates from the gas expanding from the self-crossing shock, where the gas is dense (avoiding being fully ionized) and spans a wide solid angle (providing a large emitting area). However, a caveat to keep in mind is that, during this super-Eddington phase, a large fraction of the fallback material is likely blown away by the strong radiation pressure without joining the disk.

(2) The one-zone model predicts that the outer disk collapses into a thin state due to thermal-viscous instability around $t\simeq 100\rm\, days$. The physical reason is that at this time the outer disk becomes radiatively efficient and geometrically thin while the pressure is dominated by radiation. The critical fallback rate at which the vertical collapse occurs is roughly given by (cf. eq. \ref{eq:Mdotcrit})
\begin{equation}
    \dot{M}_{\rm fb,c}\simeq {\rd \dotMEdd\over 10\rg} \simeq 0.12\, \msunyr \lrb{M\over 10^6\msun}^{1/3} \lrb{M_*\over \msun}^{-1/3} {R_*\over R_\odot} \lrb{\beta\over 2}^{-1},
\end{equation}
where we have taken a fiducial penetration parameter of $\beta=2$ for which $\rd\simeq \rT$. Since the late-time fallback rate scales as $\dotMfb(t\gtrsim t_{\rm fb})\propto t^{-5/3} M^{1/3} M_*^{1/3} R_*$ (eq. \ref{eq:fallback_rate}), we obtain an analytic estimate for the collapse time of the outer disk
\begin{equation}\label{eq:t_collapse_outer}
    t_{\rm c, outer} \simeq 80\mr{\, d} \, \lrb{\beta\over 2}^{3/5} \lrb{M_*\over \msun}^{2/5},
\end{equation}
where the normalization is numerically calibrated based on Fig. \ref{fig:evolution}. We note that the collapse time of the outer disk given by eq. (\ref{eq:t_collapse_outer}) does not depend on the BH mass.

For most TDEs with $\beta\sim \mc{O}(1)$ and $M_*\lesssim 1\msun$, the model predicts that the collapse occurs within the first few months since the earliest fallback. \citet{shen14_disk_evolution} attributes the steep X-ray flux decline in jetted TDEs at $t\simeq 1\rm\, yr$ as due to the collapse of the outer disk. Under this interpretation, these jetted TDEs are disruptions of more massive stars and/or at higher penetration parameters ($\beta\sim 10$). This is certainly possible given the low jetted TDE rate --- the Swift BAT inferred rate (after correcting for beaming) is $\lesssim1\%$ of the total TDE rate \citep{decolle20_tde_jets}. However, as shown in Fig. \ref{fig:evolution}, the X-ray lightcurves of the two optically selected (non-jetted) TDEs, AT2018fyk \citep{wevers19_at2018fyk_xray_variable, wevers21_AT2018fyk} and AT2021ehb \citep{yao22_AT2021ehb}, also show sudden flux drops by more than an order of magnitude over a few days around $t\sim 1\rm\, yr$, indicating rapid state transitions. Some other optically selected TDEs, e.g., ASASSN-14li \citep{brown17_as14li_late_time}, -15oi \citep{gezari17_as15oi_xray}, -20hx \citep{hinkle22-as20hx} show X-ray emission lasting for many hundred days without a state transition. These are in disagreement with the predictions from the one-zone model.

We speculate here that these observations can be easier explained if a fraction of the fallback gas directly lands on the innermost regions of the disk near the ISCO. This is possible because a strong self-crossing shock can cause a large spread in the angular momentum of the fallback material \citep{lu20_self_intersection}. As an example, for a flat angular momentum distribution of $\d \dotMfb/\d\ell \propto \ell^0$, we expect a fraction $\sim\!\lrb{10\rg/r_{\rm T}}^{1/2}$ of the gas to land within a radius of $10\rg$. In this case, we can estimate the collapse time of the inner disk by $\dotMfb(t_{\rm c,inner}) = \dotMEdd$. For a total fallback rate of $\dotMfb\simeq 0.5\,\msun\,\mr{yr}^{-1} (t/10^2\mr{\,d})^{-5/3} (M/10^6\msun)^{1/3} (M_*/\msun)^{4/3}$ \citep[for typical full disruptions of main-sequence stars, see Fig. 7 of][]{law-smith20_STAR_library}, we obtain the vertical collapse time for the inner disk near $10\rg$ to be
\begin{equation}\label{eq:t_collapse_inner}
    t_{\rm c,inner} \simeq 1\mr{\,yr}\, \lrb{M\over 10^6\msun}^{-1/5} \lrb{M_*\over \msun}^{3/5},
\end{equation}
which is in better agreement with observations of jetted TDEs, as well as AT2018fyk and AT2021ehb. In this picture, the disk will undergo an outside-in collapse --- a cooling front starts from the outer disk at $t_{\rm c,outer}$ and it propagates to the innermost regions at $t_{\rm c,inner}$.
If this is confirmed by future global modeling of the disk evolution, an interesting prospect is to use the timing of the steep X-ray flux drop to infer the fallback rate and hence constrain the mass of the disrupted star.

(3) Another prediction from the piecewise steady-state one-zone model with $\fsh=0$ is that the disk luminosity drops to a very low level ($\sim\! 10^{-4}L_{\rm Edd}$) after each vertical collapse. This may not be physical, since other processes may operate to limit the disk thickness above a certain value $\theta_{\rm min}$, which would prevent the accretion rate from dropping to extremely low values. For instance, on-going shock interactions between the fallback material and existing disk will maintain the pressure above the external ram pressure $\sim \!GM\rho_{\rm fb}/\rd$, where $\rho_{\rm fb}\sim \dotMfb/(\pi \rd^2 \sqrt{GM/\rd})$ is the density of the infalling gas. A fraction of the radiation generated by the circularization shocks may also heat up the disk (such that $\fsh \neq 0$). This disk heating tends to stabilize the accretion disk by raising the temperature in the gas-pressure dominated regime (see Fig. \ref{fig:boundary_fsh}). Another possibility is that the disk is supported by magnetic pressure in the low-accretion rate state \citep{begelman07_magnetic_stable, mishra20_disk_magnetic_pressure}. It is also possible that the fallback material excites spiral density waves which drives the accretion inside the disk \citep{bonnerot20_disk_formation}. Another mechanism restricting the minimum disk thickness is that, in the case of a spinning BH, Lense-Thirring precession in the Kerr spacetime causes both the fallback stream and the disk to precess around the BH's spin axis. This causes misalignment between the angular momentum vectors of different fluid elements and the misalignment will only be damped on a viscous timescale.




(4) After each vertical collapse, the accretion rate drops below the fallback rate and hence the disk mass would start to accumulate over time. When it reaches the lower-branch critical surface density $\Sigma_{\rm crit}^-$ (eq. \ref{eq:Mdotcrit}), the disk has the tendency of reviving to the thick state temporarily and this leads to sharp spikes in the accretion rate evolution. However, rapid flares from such accretion bursts (similar to dwarf novae from a cataclysmic variable system) have not been observationally confirmed. As mentioned above, it is possible that the disk thickness is maintained above a minimum value $\theta_{\rm min}$ (due to interactions with the fallback material and Lense-Thirring precession), and if the accretion rate at the lower-branch critical surface density exceeds the fallback rate, i.e., $3\pi \alpha \theta_{\rm min}^2 \omgK \rd^2 \Sigma_{\rm crit}^- > \dotMfb(\lesssim\rd)$, then the disk cannot transition back to the thick state again. Making use of $\Sigma_{\rm crit}^-$ in eq. (\ref{eq:Mdotcrit}), the criterion for avoiding a thin$\rightarrow$thick transition is
\begin{equation}
    \theta_{\rm min} > 4.6\times10^{-2} \lrb{\rd\over 10\rg}^{-7/20} \lrb{\dotMfb(\lesssim \rd)\over \dotMEdd}^{1/2}.
\end{equation}
A global disk simulation is needed to make better predictions to be compared with observations.

(5) Finally, the luminosity from the circularization shocks $L_{\rm sh}$ evolves in a smoother manner than the accretion luminosity from the disk. The small but rapid drops in $L_{\rm sh}$ along with the accretion bursts in the one-zone model (bottom panels of Fig. \ref{fig:evolution}) are due to rapid increase in the outer disk radius in the thick state. In the case where the fallback material lands onto the disk at a wide range of radii (as expected for a broad the angular momentum distribution), we expect the shock luminosity to evolve with time more smoothly than shown in Fig. \ref{fig:evolution}. The luminosity from the circularization shocks exceeds that from the accretion disk when the accretion rate is much less than the fallback rate. However, at sufficiently late time, the disk accretion power must dominate over the shock power (as a result of the rapidly declining fallback rate). TDEs often show a smooth lightcurve in the UV/optical band within the first year or two, which may be in better agreement with the circularization shock power (instead of disk accretion power).
Recently HST far-UV observations \citep{vanvelzen19_late_time_UV} indicate that the lightcurves of many TDEs flatten at such late epochs ($t\sim 5$-- 10$\rm\,yr$). In \cite{vanvelzen19_late_time_UV} and \cite{mummery20_TDE_disk}, it was proposed that the late-time flattening can be explained by a thin disk containing nearly all the fallback mass ($M_*/2$, without considering on-going fallback) and is in a state with a viscous time of decades or longer --- like in the original model of \cite{cannizzo90_disk_evolution}. However, such a model does not self-consistently explain the early time lightcurve in the first few years. Even a decade after the disruption, we still expect the disk evolution to be strongly affected by the interactions with on-going fallback gas.

\section{Summary and Future Directions}


In this review, we briefly summarize the recent progress on the disk evolution of TDEs.
The main assumption in the current modeling is that the fallback material joins the existing accretion disk at the fallback rate that is given by the stellar disruption phase. This may be a reasonable assumption provided that the circularization process is efficient as a result of strong GR apsidal precession. This allows us to make some progress by considering the qualitative behaviors of the disk on timescales much longer than the circularization time (while circumventing the highly computationally expensive simulations of the hydrodynamics of the thin fallback stream).

The approach is to model both the viscous spreading of the accretion disk and the interactions between the fallback material and the disk. The current status is that only the simplest model with both ingredients has been considered --- a piecewise steady-state one-zone model \citep{shen14_disk_evolution}. This model explains some qualitative aspects of the observed TDEs: (marked by ``$+$'' signs)


\begin{itemize}[leftmargin=1cm]
    \item[$\boldsymbol{+}$] At early time $t\lesssim100\rm\,d$, the accretion rate is comparable to the fallback rate and the emerging bolometric luminosity is mildly super-Eddington. Most of the disk mass is driven away from the system as a fast wind. If a large fraction of the bolometric luminosity is absorbed and re-emitted by a dense, optically thick gas layer located at a distance of $10^{14}$--$10^{15}\rm\, cm$ \citep{metzger16_reprocessing}, then this is capable of explaining the observed UV/optical emission near the peak luminosity. 
    \item[$\boldsymbol{+}$] As the fallback rate drops over time, the disk undergoes a thermal-viscous instability and collapses from a thick state to a thin state. This causes the accretion rate to drop by a few orders of magnitude in an abrupt way. Evidences for such a state transition are seen in jetted and non-jetted TDEs \citep{bloom11_swiftj1644_57, cenko12_swiftj2058_05, pasham15_2058jet_shut_off, wevers21_AT2018fyk, yao22_AT2021ehb}.
    \item[$\boldsymbol{+}$] After the vertical collapse, the disk accumulates in mass and could potentially go back to the thick state for a brief period of time --- this creates a limit cycle which eventually terminates at sufficiently late time when the fallback rate is insufficient to trigger an thin$\rightarrow$thick transition. AT2021ehb \citep{yao22_AT2021ehb} underwent two rapid flux drops separated by 50 days, which tentatively supports the limit cycle prediction.
\end{itemize}

However, the one-zone model is highly incomplete in that it fails to explain the following observations: (marked by ``$-$'' signs)

\begin{itemize}[leftmargin=1cm]
    \item[$\boldsymbol{-}$] The nature of the ``reprocessing layer'', which is responsible for the optical emission, is not addressed. This is because the disk wind is too fast (velocity $v_{\rm w}\sim \sqrt{GM/\rT}\simeq 0.1\rm\,c$ for typical TDEs) and hence not dense enough to absorb the X-ray and extreme-UV photons and re-emit the energy into the optical band \citep{roth16_reprocessing}.
    \item[$\boldsymbol{-}$] Many optically selected TDEs with bright X-ray emission show a delay rise of the X-ray flux by about $1\rm\, yr$. Such a delay is likely due to obscuration by Compton-thick gas along our line of sight, but the time evolution of the attenuation is not addressed by the current model (radiative transfer calculations are needed).
    \item[$\boldsymbol{-}$] The one-zone model predicts the earliest disk instability to occur at $t\simeq 80\rm\, d$ (eq. \ref{eq:t_collapse_outer}) for typical TDEs, whereas the observed cases (jetted and non-jetted TDEs) with indications of such a disk instability show that the vertical collapse occurs at much later times $t\sim 1\rm\, yr$. The reason why the model predicts an earlier collapse time is that the critical accretion rate at which the thick$\rightarrow$thin transition occurs is proportional to the disk radius $\dot{M}_{\rm crit}^+\propto \rd$ (eq. \ref{eq:Mdotcrit}), meaning that the outer disk will collapse much earlier than the inner disk. Indeed, we find that, if a fraction of the fallback material directly lands onto the inner disk, then the collapse time of the inner disk (eq. \ref{eq:t_collapse_inner}) may be in better agreement with the observed time of rapid X-ray flux drops.
    \item[$\boldsymbol{-}$] After the disk collapses into a thin state, the mass accumulation over time may cause the disk to briefly revive to the thick state. Thus, the model predicts rapid accretion flares after the first collapse. However, such flares have so far not been observed. 
    \item[$\boldsymbol{-}$] In some X-ray bright TDEs \citep[e.g., ASASSN-14li,][]{brown17_as14li_late_time}, the disk instability is not observed up to 500 days. The current model only considers the fallback as source terms in the disk mass and angular momentum conservation equations. However, in reality, the dynamical interactions between the fallback material and existing disk generates heat and drives turbulence, which may prevent the accretion disk from collapsing to a razor-thin state. Fig. \ref{fig:disksolution} shows that including shock heating ($\fsh\neq 0$) tends to stabilize the disk. It can be shown that, for sufficiently high shock heating efficiencies $\fsh$ and fallback rates, the unstable solutions disappear, meaning that the disk instability may be delayed to a later time than that in eq. (\ref{eq:t_collapse_inner}).
\end{itemize}

Future works are needed in the following directions:

\begin{itemize}[leftmargin=1cm]
    \item Long-term X-ray monitoring of nearby TDEs (selected by either optical or X-ray surveys) will (i) test if the majority of TDEs undergo the thermal-viscous instability and (ii) potentially unveil the limit-cycle behaviors (including possible missing flares).
    \item The timing of the disk state transition may only be correctly captured in a global disk model of at least 1D.
    \item The angular momentum distribution of the fallback material is likely broad (instead of a $\delta$-function at $\ell\approx \sqrt{2GM\rT}$) such that there is mass in-fall at a wide range of radii from the ISCO to a few times $\rT$. The physical cause for such a broad angular momentum distribution is most likely the self-crossing shock, which needs to be carefully modeled.
    \item Radiation hydrodynamic simulations of the interactions between the fallback material and existing accretion disk are needed to quantify the influence of on-going fallback on the accretion rate and energy-balance inside the disk.
\end{itemize}

\section*{Acknowledgments}
I would like to thank Yuhan Yao and Thomas Wevers for sharing their X-ray data files, Bin-Bin Zhang for the invitation to write this review. I also thank Cl\'{e}ment Bonnerot for collaboration. WL was supported by the Lyman Spitzer, Jr. Fellowship at Princeton University.

{\small
\bibliographystyle{mnras}
\bibliography{refs}
}

\end{document}